# Stress-induced rearrangements of cellular networks: consequences for protection and drug design

Máté S. Szalay[a,¶], István A. Kovács [a,¶], Tamás Korcsmáros[a,¶], Csaba Böde [b,¶] and Péter Csermely [a,*]

*Department of [a]Medical Chemistry and [b]Biophysics and Radiation Biology,
Semmelweis University, Puskin str. 9, H-1088 Budapest, Hungary*

**Abstract.** The complexity of the cells can be described and understood by a number of networks such as protein-protein interaction, cytoskeletal, organelle, signalling, gene transcription and metabolic networks. All these networks are highly dynamic producing continuous rearrangements in their links, hubs, network-skeleton and modules. Here we describe the adaptation of cellular networks after various forms of stress causing perturbations, congestions and network damage. Chronic stress decreases link-density, decouples or even quarantines modules, and induces an increased competition between network hubs and bridges. Extremely long or strong stress may induce a topological phase transition in the respective cellular networks, which switches the cell to a completely different mode of cellular function. We summarize our initial knowledge on network restoration after stress including the role of molecular chaperones in this process. Finally, we discuss the implications of stress-induced network rearrangements in diseases and ageing, and propose therapeutic approaches both to increase the robustness and help the repair of cellular networks.

**1. Introduction: Cellular networks, stress responses, adaptation and learning**
The complexity of the cells can be described reasonably well, if we catalogue those interactions of cellular molecules only, which have a relatively high affinity, and, therefore, are unique and specific interactions of the cell. Here the interacting molecules behave as network elements, and their interactions form the weighted, but not necessarily directed links of the respective structural network. Alternatively, we may also envision directed links as representations of signalling or metabolic processes of the functional networks in the cell (Table 1. [1-3]).

Cellular networks often form small worlds, where two elements of the network are separated by only a few other elements. Networks of our cells contain hubs, i.e. elements, which have a large number of neighbours. These networks can be dissected to overlapping modules, which form hierarchical communities [4-6]. However, this summary of the major features of cellular networks is largely a generalization, and needs to be validated through critical scrutiny of the datasets, sampling procedures and methods of data analysis at each network examined [7,8].

The word, "stress" has been coined by Hans Selye [9,10], who was born a hundred years before the writing and publication of this paper. Here we use a definition of stress from the point of the cellular networks. Stress is any unexpected, large and sudden perturbation of the cellular network, to which the network (1) does not have a prepared adaptive response or (2) does not have enough time to mobilize the adaptive response. From the network point of view we talk about an adaptive response, if a massive network rearrangement occurs. Learning of networks can be differentiated from adaptation, if we restrict the learning to those network rearrangements, which are extended only to a few links and network elements.

---

[*]Máté S. Szalay (szalaymate@gmail.com), István A. Kovács (steve3281@bolyai1.elte.hu), Tamás Korcsmáros (korcsmaros@gmail.com) and Csaba Böde (csabi@puskin.sote.hu) started their research as members of the Hungarian Research Student Association (www.kutdiak.hu), which provides research opportunities for talented high school students since 1996.
*Correspondence to:* Péter Csermely, Department of Medical Chemistry, Semmelweis University School of Medicine, Budapest, P.O. Box 260. H-1444 Hungary. Telephone: +361-266-2755 extn.: 4102. fax: +361-266-6550. E-mail: csermely@puskin.sote.hu

The cellular response to stress involves a number of specific signalling events as well as the activation and extensive synthesis of molecular chaperones, many of which are also called heat shock, or stress proteins [11-13]. While the signalling events mostly prepare for the specific cellular adaptation steps in the metabolism, membranes, cytoskeleton, and for those of other cellular elements and functions, chaperones provide a general response to stress by repairing damaged proteins (and RNA-s).

## 2. Stress of networks

Stress can be a single network perturbation (which becomes noteworthy, if it was large enough), but it is often repeated, which may cause a congestion of the perturbations at special points of the cellular networks (Fig. 1). Congestion often affects most the communication boundaries, such as the central hubs of hierarchical networks, or the overlaps of network modules [14]. Rather paradoxically, the chances to develop congestion become higher, if the network is denser, meaning that the average number of neighbours is higher in the network [15].

If perturbations repeatedly arrive within the relaxation time of the element or group of elements, network damage may occur. Network may also be damaged, if the propagation of such a single but large perturbation becomes prevented. As a very simple scenario, extensively repeated smaller conformational changes of proteins (or an extremely large change in the shape of a protein) may induce a partial unfolding, which is large enough to misfold, and denature the protein in question.

Continuing the description of the above scenario, altered shapes of proteins will not properly serve the very same contacts the respective protein had before. In other words, links to the former neighbours in the protein-protein interaction network will be more transient, weaker. Consequently, stress induces a decrease in the average link-strength and link-density. Deleting links may be actually beneficial, and may help to prevent the propagation of damage [16]. Stress-induced shift from strong links to weaker ones may actually form a part of a 'Le-Chatelier type network principle' meaning that upon disturbing the former network equilibrium, the network starts to have an automatic attenuation of link-strength, which acts a fuse, re-channels the perturbations to alternative routes, and induces an extra stabilization counteracting the original damage (Fig. 1. [3,17]).

## 3. Network rearrangements in stress

Stress-induced decrease in the strength and number of links leads to a gradual detachment of network elements from each other. This results in a larger number of 'lonely' elements, but – more importantly – induces an increased competition between the hubs of the network for the remaining links. During a prolonged stress the segregation of 'looser hubs' and 'winner hubs' will occur, where winner hubs will be preferentially those, which are more flexible, can endure the transmission of perturbations better, had more links to their neighbours before, and the average strength of these links was stronger. Similarly, stress-induced scarcity of stronger links also leads to an increased competition of bridges, resulting in weaker, more rigid 'looser bridges' and 'winner bridges' (Fig. 2.). Extending this description winner hubs and bridges may have a larger repertoire of game strategies in the iterative scenarios of making and maintaining protein-protein interactions, which might also be called protein games [18].

Parallel with the stress-induced segregation of looser and winner hubs and bridges, stress also provokes an increased de-coupling of network modules. The overlap decreases between modules (Fig 2.). This leads to simpler, less regulated, more specialized cellular functions. This fits well to the requirement for more efficient energy utilization and preparation for larger autonomy of various cellular modules during stress. (A larger autonomy is expected to occur, if there is an increased chance for a damage of various parts of the network – which is exactly the scenario during stress, where random-type damages may hit one or another module.)

The larger autonomy may be increased to such an extent that certain modules become completely de-coupled from the rest of the network. This quarantining may affect those modules which suffered the largest damage or may isolate those segments, which would utilize the most resources from the diminished reserves.

Modular rearrangements certainly affect the hierarchical structure of cellular modules [19]. Elementary modules of some super-modules may coalesce into a single grand-module and suffer a significant reduction, while other super-modules may fall apart to a loose or even isolated assembly of single modules.

All these 'simplifications' of cellular networks during stress resemble to an accelerated and reversible version of the reductive evolution of symbiotic organisms, especially, where the engulfment by the host provides a safe and stable environment for the 'guest', e.g. a parasite [20,21]. In both processes major segments of the original networks become attenuated parallel with a specialization of the network structure for a specific set of environmental conditions provided by either the stress or the host. This network simplification gives a more rigid structure, where most of the original universal and flexible adaptation strategies were temporarily or irreversibly lost.

During stress, chaperones become increasingly occupied by damaged proteins and a so-called 'chaperone overload' occurs [22]. Since chaperones often couple various cellular modules of protein-protein interaction [23] and cellular organelle networks [24], their inhibition might lead to a de-coupling of all these chaperone-mediated contacts between network modules providing an additional safety measure for the cell [13]. (De-coupling of modules may stop the propagation of network damage at the modular boundaries.)

In chronic stress or extreme changes in the environment, smaller or larger parts of cellular network may undergo a topological phase transition [25]. When the cell has plenty of resources, there is ample energy to build a large amount of links between network elements, which means that the resulting topology will be very close to that of a random network. If the network experiences stress, and if its resources become more and more depleted, in agreement with the above notions, a larger and larger number of links will cease to operate. This leads to a discrimination of network elements. A few of them will retain and even gain links, while most of them will be pauperized. The 'link-winners' will become the hubs of the novel network structure. If the stress is stronger or it lasts longer, an increased competition of hubs will occur, and in an extreme case, the network may be switched to a star-network, where the 'winner hub takes all' and an extremely centralized, highly hierarchical structure develops. If the resources will become even smaller, the star-network collapses and a number of isolated, small groups will be formed. This corresponds to the death of the former gross structure, which we called cell or organism (Fig. 2). The latter, disintegration-type topological phase transition may be preceded by quarantining the most damaged parts of the network, and might accompany various forms of programmed cell death [26].

**4. Network re-establishment after stress**
When the stress is over, and cellular resources slowly start to get back to normal again, cellular networks start to re-establish those links, which were ceased to operate during stress. Bridges, local hubs are re-built, modules are re-coupled. As a gross summary of these processes, the cell re-establishes its lost repertoire of weak links, which enable its networks to a large number of flexible changes. In this way the re-gaining of the links shed during stress can be envisioned as a purchase of a general 'insurance', which enables the stressed cell to recover from its former, rigid state highly specialized to the given form of stress, and to attain a more flexible structure, which will be able cope with a large number of unexpected changes in the future.

These processes are helped by the newly synthesized molecular chaperones, since their low affinity interactions effectively sample a large number of proteins, and allow the re-arrangement of hubs, re-formation of bridges and binding of de-coupled modules each other in a very flexible, partially stochastic manner. Thus, chaperones give the cell a refined and flexible way for the gradual build-up of the complex modular structure and function, when the stress is already over [23,27].

## 5. Stressed cellular networks in disease and ageing

The increased vulnerability of stressed cellular networks, together with the significantly reduced range of adaptive responses of stressed networks become especially critical, if the cellular network already experienced a previous damage. This happens under the repeated waves of chronic stress, but also prevalent in disease and in aged cells. In these scenarios stress-induced severing of various links in cellular networks affects a network structure, which has been already weakened by the preceding permanent damage. An additional type of danger is raised by the fact that in disease and ageing the noise level is already much higher than usual [28]. If disease and age-induced noise is accompanied by the extra, stress-generated noise, it may well go beyond the tolerable threshold, and induces an 'error-catastrophe'.

## 6. Robustness and repair of stressed networks: consequences to pharmacology and drug design

Networks have a number of efficient measures to protect themselves from stress-induced damage. If the traffic of the most central network elements is redistributed to other, non-central nodes, the network capacity can be increased by ten times [29,30]. If a network element is destroyed, an isoform may be synthesized as a back-up, or an emergency rewiring of its neighbors can also save the network [31]. Additionally, the selective removal of network elements and links with either a small load, or a large excess of overload also diminishes the size of the cascading damage [16].

Efficient repair of the multiple rearrangements and defects of disease-, ageing- and stress-affected cellular networks is rather seldom reached by a single-target drug having a well-designed, high affinity interaction with one of the cellular proteins [32]. In agreement with this general assumption, several examples show that multi-target therapy may be superior to the usual single-target approach. The best known examples of multi-target drugs include Aspirin, Metformin or Gleevec as well as combinatorial therapy and natural remedies, such as herbal teas [33]. Due to the multiple regulatory roles of chaperones, chaperone-modulators provide additional examples for multi-target drugs. Indeed, chaperone substitution (in the form of chemical chaperones), the help of chaperone induction and chaperone inhibition are all promising therapeutic strategies [34-37].

## 7. Summary and perspectives

Recent progress in network science and, especially, our emerging knowledge of network dynamics provides a unique chance to understand and modulate stress-induced changes in cellular networks. We have an initial idea on stress-induced network rearrangements as well as on a few mechanisms helping the cell to preserve robustness under these conditions. However, a number of key issues have not been tackled yet both from the theoretical and the experimental points of view.

- We do not have a detailed view on the vulnerable points of real, cellular networks suffering from congestions or being exposed to damage during stress.
- We are at the very beginning to understand stress-induced network rearrangements: the exploration of the rules of hub- and bridge-competition as well as changes in the hierarchical modular structure is largely missing.
- Parallel data-sets showing the differences between protein-protein interaction, organelle and functional cellular networks under normal and stressful conditions are missing.
- The exploration of topological phase transitions of cellular networks by comparing their topology in extremely resource-rich and resource-poor environments awaits experimentation.
- Our knowledge on the re-establishment, re-building of cellular networks after stress is practically zero.
- We need a much better understanding of cellular network changes in disease and ageing.
- Lastly, but perhaps most importantly, besides chaperones, and chaperone-related therapies we do not have a detailed knowledge on the mechanisms helping cellular networks to cope with their stress and on the possibilities for efficient therapeutic interventions.

We are quite certain that stressed networks will give a lot of excitement and pleasure for systems biologists, who would like to understand the dynamics of cellular networks. As a result of these studies the emergence of network-based therapies is expected, where the target-sets of multi-target drugs will be identified using our knowledge on the vulnerable points (hot-spots) of cellular networks in stress, disease and ageing.

**Acknowledgments**
The authors would like to thank members of the LINK-group (www.weaklinks.sote.hu) for helpful discussions. Work in the authors' laboratory was supported by research grants from the EU (FP6-506850, FP6-016003) and by the Hungarian National Research Initiative (NKFP-1A/056/2004 and KKK-0015/3.0).

**References**
[1] Barabasi, A.L. and Oltvai, Z.N. (2004) Network biology: understanding the cell's functional organization. Nat. Rev. Genet. 5, 101–113.
[2] Boccaletti, S., Latora, V., Moreno, Y., Chavez, M. and Hwang, D.-U. (2006) Complex networks: structure and dynamics. Physics Rep. 424, 175–308.
[3] Csermely, P. (2006) Weak links: a universal key for network diversity and stability, Springer Verlag, Heidelberg.
[4] Watts, D.J. and Strogatz, S.H. (1998) Collective dynamics of 'small-world' networks. Nature 393, 440–442.
[5] Barabasi, A.L. and Albert, R. (1999) Emergence of scaling in random networks. Science 286, 509–512.
[6] Palla, G., Derenyi, I., Farkas, T. and Vicsek, T. (2005) Uncovering the overlapping community structure of complex networks in nature and society. Nature 435, 814–818.
[7] Arita, M. (2004) The metabolic world of *Escherichia coli* is not small. Proc. Natl. Acad. Sci. USA 101, 1543–1547.
[8] Tanaka, R., Yi, T. M. and Doyle, J. (2005) Some protein interaction data do not exhibit power law statistics; FEBS Lett. 579, 5140–5144.
[9] Selye, H. (1955) Stress and disease. Science 122, 625–631.
[10] Selye, H. (1956) The stress of life. McGraw-Hill, New York NY, USA.
[11] Hartl, F.-U. (1996) Molecular chaperones in cellular protein folding. Nature 381, 571–580.
[12] Bukau, B. and Horwich, A.L. (1998) The Hsp70 and Hsp60 chaperone machines. Cell 92, 351–366.
[13] Soti, C., Pal, C., Papp, B. and Csermely, P. (2005) Chaperones as regulatory elements of cellular networks. Curr. Op. Cell Biol. 17, 210–215.
[14] Gfeller, D., Chappelier, J.-C. and De Los Rios, P. (2005) Finding instabilities in the community structure of complex networks. Phys. Rev. E 72, 056135.
[15] Park, K., Lai, Y.-C., Zhao, L. and Ye, N. (2005) Jamming in complex gradient networks. Phys. Rev. E 71, 065105.
[16] Motter, A.E. (2004) Cascade control and defense in complex networks. Phys. Rev. Lett. 93, 098701.
[17] Csermely, P. (2004) Strong links are important, but weak links stabilize them. Trends Biochem. Sci. 29, 331–334.
[18] Kovacs, I.A., Szalay, M.S. and Csermely, P. (2005) Water and molecular chaperones act as weak links of protein folding networks: energy landscape and punctuated equilibrium changes point towards a game theory of proteins. FEBS Lett. 579, 2254-2260.
[19] Ravasz, R., Somera, A.L., Mongru, D.A., Oltvai, Z.N. and Barabasi, A.L. (2002) Hierarchical organization of modularity in metabolic networks. Science 297, 1551-1555.
[20] Papp, B., Pal, C. and Hurst L.D. (2004) Metabolic network analysis of the causes and evolution of enzyme dispensability in yeast. Nature 429, 661–664.
[21] Pál, C., Papp, B., Lercher, M.J., Csermely, P., Oliver, S.G. and Hurst, L.D. (2006) Chance and necessity in the evolution of minimal metabolic networks. Nature 440, 667–670.
[22] Csermely P. (2001) Chaperone-overload as a possible contributor to "civilization diseases": atherosclerosis, cancer, diabetes. Trends Genet 17, 701–704.
[23] Korcsmaros, T., Kovacs, I.A., Szalay, M.S. and Csermely, P. (2007) Molecular chaperones: the modular evolution of cellular networks. J. Biosci. in press (www.arxiv.org/q-bio.MN/0701030).
[24] Szabadkai, G., Bianchi, K., Varnai, P., De Stefani, D., Wieckowski, M.R., Cavagna, D., Nagy, A.I., Balla, T. and Rizzuto, R. (2006) Chaperone-mediated coupling of endoplasmic reticulum and mitochondrial Ca2+ channels. J. Cell Biol. 175, 901–911.
[25] Derenyi, I., Farkas, I., Palla, G. and Vicsek, T. (2004) Topological phase transitions of random networks. Physica A 334, 583–590.
[26] Soti, C., Sreedhar, A.S. and Csermely, P. (2003) Apoptosis, necrosis and cellular senescence: chaperone occupancy as a potential switch. Ageing Cell 2, 39–45.
[27] Lewandowska, A., Gierszewska, M., Marszalek, J. and Liberek. K. (2006) Hsp78 chaperone functions in restoration of mitochondrial network following heat stress. Biochim. Biophys. Acta 1763, 141–151.
[28] Goldberger, A.L., Amaral, L.A.N., Hausdorf, J.M., Ivanov, P.C., Peng, C.-K. and Stanley, H.E. (2002) Fractal dynamics in physiology: alterations with disease and ageing. Proc. Natl. Acad. Sci. U. S. A. 99, 2466–2472.
[29] Ghim, C-M., Oh, E., Goh, K.-I., Khang, B. and Kim, D. (2004) Packet transport along the shortest pathways in scale-free networks. Eur. Phys. J. B 38, 193–199.
[30] Yan, G., Zhou, T., Hu, B., Fu, Z.-Q. and Wang, B.-H. (2006) Efficient routing on complex networks. Phys. Rev. E 73, 046108.
[31] Hayashi, Y. and Miyazaki, T. (2005) Emergent rewirings for cascades on correlated networks. www.arxiv.org/cond-mat/0503615.

[32] Kitano, H. (2004) Biological robustness. Nature Rev. Genetics 5, 826–837.
[33] Csermely, P., Agoston, V. and Pongor, S. (2005) The efficiency of multi-target drugs: the network approach might help drug design. Trends Pharmacol. Sci. 26, 178–182.
[34] Bernier, V., Lagace, M., Bichet, D.G and Bouvier, M. (2004) Pharmacological chaperones: potential treatment for conformational diseases. Trends Endocrinol. Metab. 15, 222–228.
[35] Neckers, L. and Neckers, K. (2005) Heat-shock protein 90 inhibitors as novel cancer chemotherapeutics - an update. Expert Opin. Emerg. Drugs 10, 137–149.
[36] Vigh, L., Literati, P.N., Horvath, I., Torok, Z., Balogh, G., Glatz, A., Kovacs, E., Boros, I., Ferdinandy, P., Farkas, B., Jaszlits, L., Jednakovits, A., Koranyi, L. and Maresca, B. (1997) Bimoclomol: a nontoxic, hydroxylamine derivative with stress protein-inducing activity and cytoprotective effects. Nat. Med. 3, 1150–1154.
[37] Soti, C., Nagy, E., Giricz, Z., Vigh, L., Csermely, P. and Ferdinandy, P. (2005) Heat shock proteins as emerging therapeutic targets. Br. J. Pharmacol. 146, 769–780.

**Table 1**
**Cellular networks**

| Name of cellular network | Network elements | Network links |
| --- | --- | --- |
| Protein interaction network | Cellular proteins | Transient or permanent bonds |
| Cytoskeletal network | Cytoskeletal filaments | Transient or permanent bonds |
| Organelle network | Membrane segments (membrane vesicles, domains, rafts, of cellular membranes) and cellular organelles (mitochondria, lysosomes, segments of the endoplasmic reticulum, etc.) | Proteins, protein complexes and/or membrane vesicles, channels |
| Signalling network | Proteins, protein complexes, RNA (such as micro-RNA) | Highly specific interactions undergoing a profound change (either activation or inhibition), when a specific signal reaches the cell |
| Metabolic network | Metabolites, small molecules, such as glucose, or adenine, etc. | Enzyme reactions transforming one metabolite to the other |
| Gene transcription network | Transcriptional factors or their complexes and DNA gene sequences | Functional (and physical) interactions between transcription factor proteins (sometimes RNA-s) and various parts of the gene sequences in the cellular DNA |

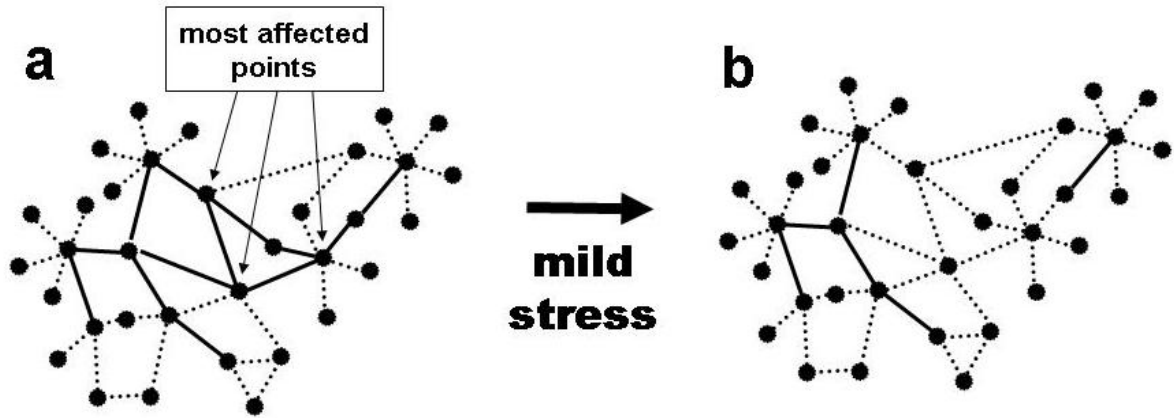

**Fig. 1. Network perturbations, congestions and damage: a 'Le-Chatelier type principle' of network stabilization after stress.** (a) The congestion of network perturbations is preferentially observed at communication boundaries, such as central hubs of hierarchical networks or overlaps of network modules. (b) Extensively repeated or large perturbations may lead to the damage of most affected network elements, which makes the links of the given element weaker to its neighbours as before. Damage-induced link-weakening may act as a fuse and by re-channelling the perturbation to alternative routes of weak links may counteract stress-induced network destabilization.

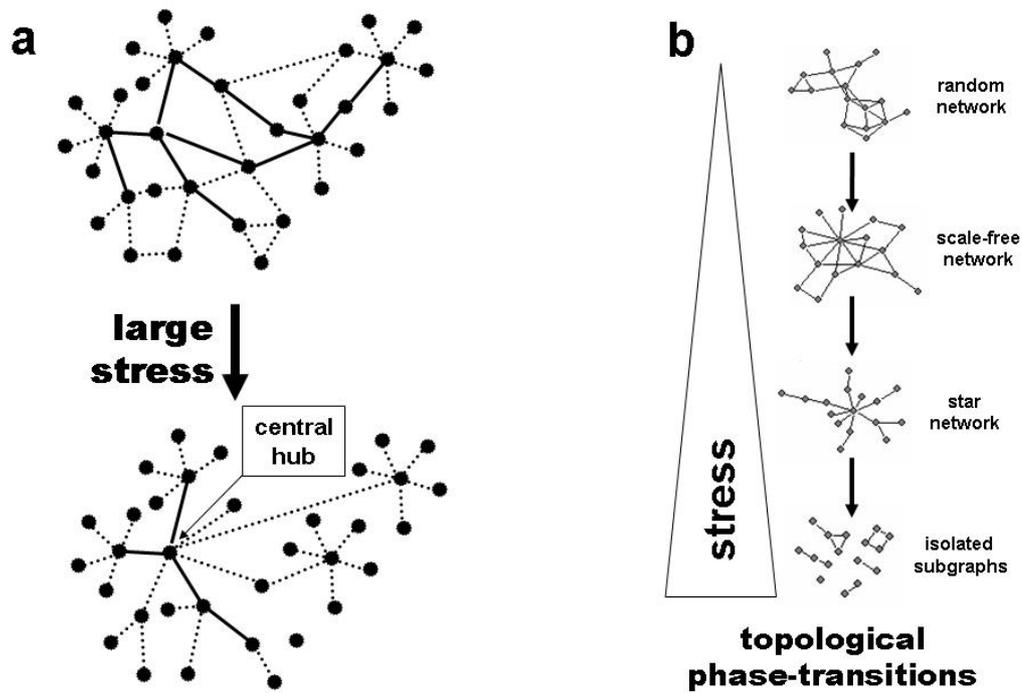

**Fig. 2. Network rearrangements in stress.** (a) Stress-induced decrease in the strength and number of links leads to detached elements, and results in an increased competition between the strongest hubs and bridges for remaining links. Parallel with this, an increased de-coupling (in extreme case: quarantining) of network modules is observed, which leads to simpler, less regulated, more specialized cellular functions. (b) In chronic stress or extreme changes in the environment parts of cellular network may undergo a topological phase transition, where the distribution of the number of neighbours becomes more and more uneven. Here the topology changes from a random network to a hub-containing network, where the number of neighbours has a scale-free distribution, then a star-network develops, where a 'dictator-hub' attains most connections, and finally the network falls apart to densely-connected small groups – called isolated subgraphs.